\documentclass[a4paper]{article}

\usepackage{INTERSPEECH2019}
\usepackage{multirow}
\usepackage{makecell}
\usepackage[flushleft]{threeparttable}
\usepackage{hyperref}

\title{Jasper: An End-to-End Convolutional Neural Acoustic Model}
\name{
Jason Li$^1$, Vitaly Lavrukhin$^1$, Boris Ginsburg$^1$, Ryan Leary$^1$, Oleksii Kuchaiev$^1$, \\
Jonathan M. Cohen$^1$, Huyen Nguyen$^1$, Ravi Teja Gadde$^2$\thanks{$^2$ Work was conducted while the author was at NVIDIA}
}

\address{
  $^1$NVIDIA, Santa Clara, USA \\
  $^2$New York University, New York, USA
}
\email{\{jasoli,vlavrukhin,bginsburg,rleary,okuchaiev,jocohen,chipn\}@nvidia.com,rtg267@nyu.edu
} 
\begin{document}

\maketitle

\begin{abstract}
In this paper we report state-of-the-art results on LibriSpeech among end-to-end speech recognition models without any external training data. Our model, Jasper, uses only 1D convolutions, batch normalization, ReLU, dropout, and residual connections. To improve training, we further introduce a new layer-wise optimizer called NovoGrad.
Through experiments, we demonstrate that the proposed deep architecture performs as well or better than more complex choices. Our deepest Jasper variant uses 54 convolutional layers. With this architecture, we achieve 2.95\% WER using a beam-search decoder with an external neural language model and 3.86\% WER with a greedy decoder on LibriSpeech test-clean. We also report competitive results on Wall Street Journal and the Hub5'00 conversational evaluation datasets.
\end{abstract}

\section{Introduction}
Conventional automatic speech recognition (ASR) systems typically consist of several independently learned components: an acoustic model to predict context-dependent sub-phoneme states (senones) from audio, a graph structure to map senones to phonemes, and a pronunciation model to map phonemes to words. Hybrid systems combine hidden Markov models to model state dependencies with neural networks to predict states \cite{Waibel1989, Bengio1992, Graves2005, Hinton2012}. Newer approaches such as end-to-end (E2E) systems reduce the overall complexity of the final system.

Our research builds on prior work that has explored using time-delay neural networks (TDNN), other forms of convolutional neural networks, and Connectionist Temporal Classification (CTC) loss \cite{graves2006, Zhang2016, collobert2016}. We took inspiration from wav2letter \cite{collobert2016}, which uses 1D-convolution layers. Liptchinsky et al. \cite{liptchinsky2017based} improved wav2letter by increasing the model depth to 19 convolutional layers and adding Gated Linear Units (GLU) \cite{Dauphin2017GLU}, weight normalization \cite{Salimans2016WeightNorm} and dropout.

By building a deeper and larger capacity network, we aim to demonstrate that we can match or outperform non end-to-end models on the LibriSpeech and 2000hr Fisher+Switchboard tasks. Like wav2letter, our architecture, Jasper, uses a stack of 1D-convolution layers, but with ReLU and batch normalization \cite{IoffeS15BatchNorm}. We find that ReLU and batch normalization outperform other activation and normalization schemes that we tested for convolutional ASR. As a result, Jasper's architecture contains only 1D convolution, batch normalization, ReLU, and dropout layers -- operators highly optimized for training and inference on GPUs.

It is possible to increase the capacity of the Jasper model by stacking these operations. Our largest version uses 54 convolutional layers (333M parameters), while our smaller model uses 34 (201M parameters). We use residual connections to enable this level of depth. We investigate a number of residual options and propose a new residual connection topology we call \textit{Dense Residual (DR)}.

Integrating our best acoustic model with a Transformer-XL \cite{dai2018transformer} language model allows us to obtain new state-of-the-art (SOTA) results on LibriSpeech \cite{panayotov2015librispeech} test-clean of 2.95\% WER and SOTA results among end-to-end models\footnote{We follow Hadian et. al's definition of end-to-end \cite{Hadian2018}: ``flat-start training of a single DNN in one stage without using any previously trained models, forced alignments, or building state-tying decision trees.''} on LibriSpeech test-other. We show competitive results on Wall Street Journal (WSJ), and 2000hr Fisher+Switchboard (F+S).  Using only greedy decoding without a language model we achieve 3.86\% WER on LibriSpeech test-clean.

This paper makes the following contributions:
\begin{enumerate}
    \item We present a computationally efficient end-to-end convolutional neural network acoustic model.
    \item We show ReLU and batch norm outperform other combinations for regularization and normalization, and residual connections are necessary for training to converge.
    \item We introduce \textit{NovoGrad}, a variant of the Adam optimizer~\cite{kingma} with a smaller memory footprint.
    \item We improve the SOTA WER on LibriSpeech test-clean.
\end{enumerate}

\section{Jasper Architecture}

Jasper is a family of end-to-end ASR models that replace acoustic and pronunciation models with a convolutional neural network. Jasper uses mel-filterbank features calculated from 20ms windows with a 10ms overlap, and outputs a probability distribution over characters per frame\footnote{We use 40 features for WSJ and 64 for LibriSpeech and F+S.}. Jasper has a block architecture: a Jasper $B$x$R$ model has $B$ blocks, each with $R$ sub-blocks. Each sub-block applies the following operations: a 1D-convolution, batch norm, ReLU, and dropout. All sub-blocks in a block have the same number of output channels.

Each block input is connected directly into the last sub-block via a residual connection.  The residual connection is first projected through a 1x1 convolution to account for different numbers of input and output channels, then through a batch norm layer. The output of this batch norm layer is added to the output of the batch norm layer in the last sub-block. The result of this sum is passed through the activation function and dropout to produce the output of the current block. 

The sub-block architecture of Jasper was designed to facilitate fast GPU inference. Each sub-block can be fused into a single GPU kernel: dropout is not used at inference-time and is eliminated, batch norm can be fused with the preceding convolution, ReLU clamps the result, and residual summation can be treated as a modified bias term in this fused operation.

\begin{figure}[t]
  \centering
  \includegraphics[width=\linewidth]{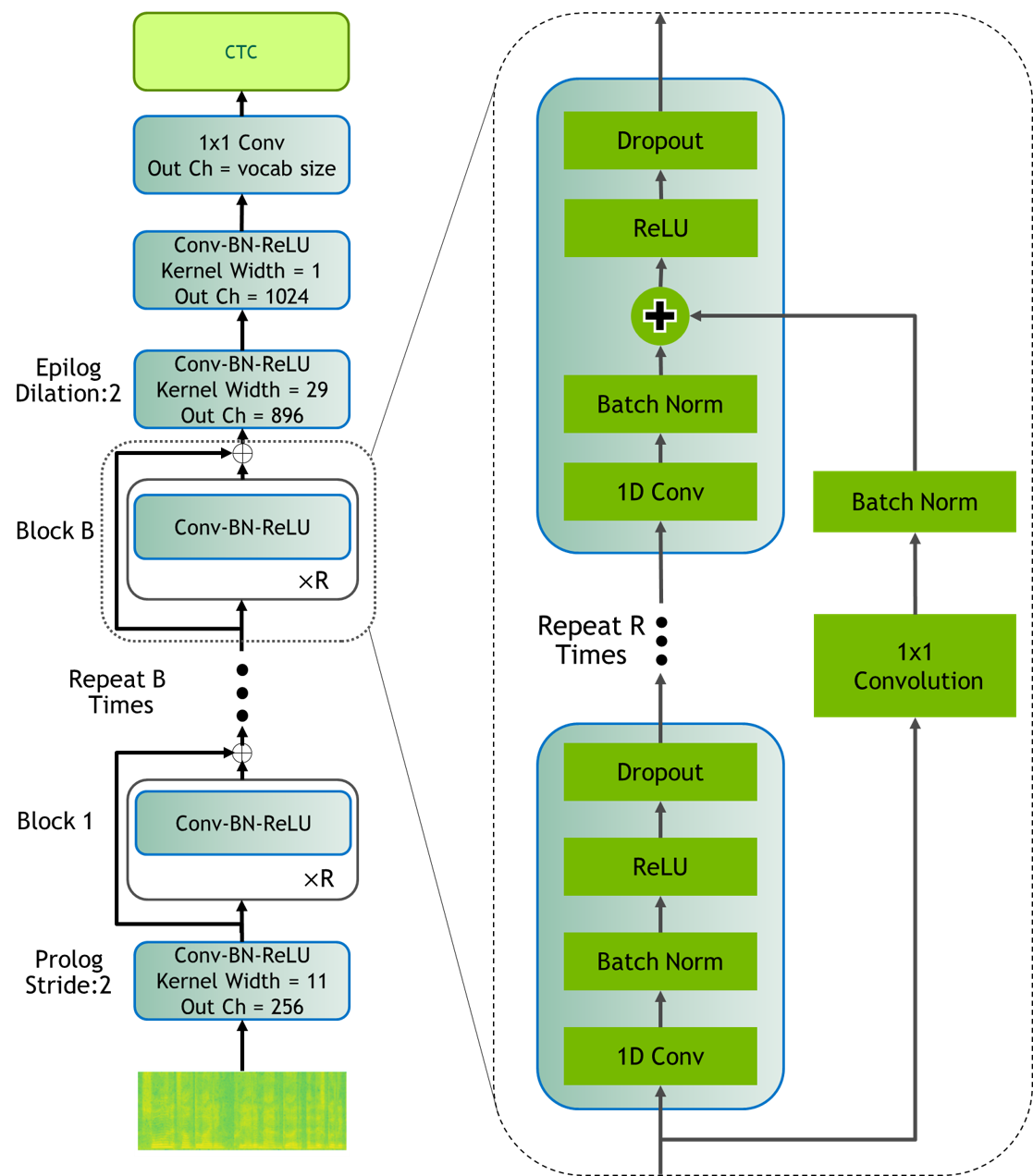}
  \caption{Jasper $B$x$R$ model: $B$ - number of blocks, $R$ - number of sub-blocks.}
  \label{fig:jasper_arch}
\end{figure}

All Jasper models have four additional convolutional blocks: one pre-processing and three post-processing. See Figure~\ref{fig:jasper_arch} and Table \ref{tab:JasperParams} for details.

{\renewcommand{\arraystretch}{1.1}
\begin{table}[!h]
\caption{Jasper 10x5: 10 blocks, each consisting of 5 1D-convolutional sub-blocks, plus 4 additional blocks.
}

\label{tab:JasperParams}
\centering
\scalebox{0.8}{
\begin{tabular}{c c c c c c} 
 \toprule
  \textbf {\# Blocks} & \textbf{Block} & \textbf{Kernel} & \textbf{\thead{\# Output\\Channels}} & \textbf{Dropout} & \textbf{\thead{\# Sub\\Blocks}} \\
 \midrule
 \multirow{2}{*}{1} & \multirow{2}{*}{Conv1} & \makecell[t]{%
 11\\%
 \textit{stride=2}} & \multirow{2}{*}{256} & \multirow{2}{*}{0.2} & \multirow{2}{*}{1}\\
 2 & B1 & 11 & 256 & 0.2 & 5 \\
 2 & B2 & 13 & 384 & 0.2 & 5 \\
 2 & B3 & 17 & 512 & 0.2 & 5 \\
 2 & B4 & 21 & 640 & 0.3 & 5 \\ 
 2 & B5 & 25 & 768 & 0.3 & 5 \\
 \multirow{2}{*}{1} & \multirow{2}{*}{Conv2} & \makecell[t]{%
 29\\%
 \textit{dilation=2}} & \multirow{2}{*}{896} & \multirow{2}{*}{0.4} & \multirow{2}{*}{1} \\
 1 & Conv3 & 1 & 1024  & 0.4 & 1 \\
 1 & Conv4 & 1 & \# graphemes & 0 & 1 \\
 \bottomrule
\end{tabular}
}
\end{table}
}

We also build a variant of Jasper, \textit{Jasper Dense Residual} (DR). Jasper DR follows DenseNet \cite{huang2016} and DenseRNet \cite{Tang2018}, but instead of having dense connections within a block, the output of a convolution block is added to the inputs of all the following blocks. While DenseNet and DenseRNet concatenates the outputs of different layers, Jasper DR adds them in the same way that residuals are added in ResNet. As explained below, we find addition to be as effective as concatenation.

\begin{figure}[t]
  \centering
  \includegraphics[width=\linewidth]{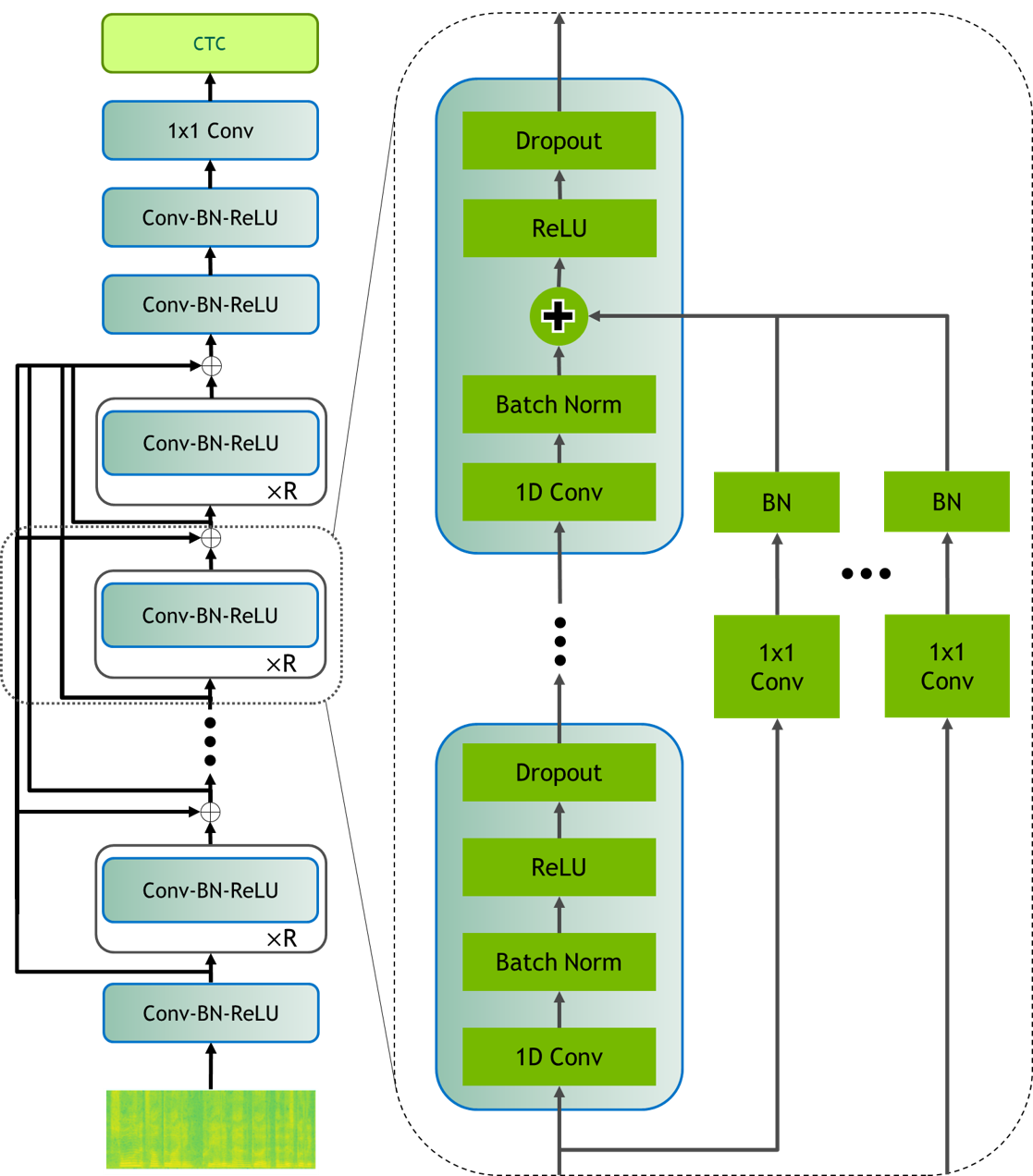}
  \caption{Jasper Dense Residual}
  \label{fig:jasper_dense}
\end{figure}

\subsection{Normalization and Activation}
In our study, we evaluate performance of models with:
\begin{itemize}
    \item 3 types of normalization: batch norm \cite{IoffeS15BatchNorm}, weight norm \cite{ Salimans2016WeightNorm}, and layer norm  \cite{Ba2016LayerNorm}
    \item 3 types of rectified linear units: ReLU, clipped ReLU (cReLU), and leaky ReLU (lReLU) 
    \item 2 types of gated units: gated linear units (GLU) \cite{Dauphin2017GLU}, and gated activation units (GAU) \cite{Oord2016PixelCNN}
\end{itemize}
All experiment results are shown in Table \ref{tab:NormAct}. We first experimented with a smaller Jasper5x3\footnote{
Jasper 5x3 models contain one block of each B1 to B5.
}
model to pick the top 3 settings before training on larger Jasper models. We found that layer norm with GAU performed the best on the smaller model. Layer norm with ReLU and batch norm with ReLU came second and third in our tests. Using these 3, we conducted further experiments on a larger Jasper10x4. For larger models, we noticed that batch norm with ReLU outperformed other choices. Thus, leading us to decide on batch normalization and ReLU for our architecture.

During batching, all sequences are padded to match the longest sequence. These padded values caused issues when using layer norm. We applied a sequence mask to exclude padding values from the mean and variance calculation. Further, we computed mean and variance over both the time dimension and channels similar to the sequence-wise normalization proposed by Laurent et al. \cite{LaurentSeqWiseBN}. In addition to masking layer norm, we additionally applied masking prior to the convolution operation, and masking the mean and variance calculations in batch norm. These results are shown in Table~\ref{tab:SeqMask}. Interestingly, we found that while masking before convolution gives a lower WER, using masks for both convolutions and batch norm results in worse performance.

As a final note, we found that training with weight norm was very unstable leading to exploding activations.

\begin{table}[th]
\centering
\caption{Normalization and Activation: Greedy WER, LibriSpeech after 50 epochs}
\label{tab:NormAct}
\scalebox{0.8}{
%\begin{tabular}{|c|c|c|c|} 
\begin{tabular} {c c c c c}
 \toprule
     \multirow{2}{*}{\textbf{Model}} &
     \multirow{2}{*}{\textbf{Normalization}} & \multirow{2}{*}{\textbf{Activation}} & \multicolumn{2}{c}{\textbf{Dev}} \\
 & & & \textbf{Clean} & \textbf{Other} \\
 \midrule
 \multirow{15}{*}{Jasper 5x3} & \multirow{5}{*}{Batch Norm} & ReLU & 8.82 & 23.26 \\
 & & cReLU & 8.89 & 23.02 \\
 & & lReLU & 11.31 & 26.90 \\
 & & GLU & 9.46 & 24.30 \\
 & & GAU & 9.41 & 24.65 \\
 \cmidrule{2-5}
 & \multirow{5}{*}{Layer Norm} & ReLU & 8.82 & \textbf{22.83} \\
 & & cReLU & 9.14 & 23.26 \\
 & & lReLU & 11.29 & 26.35 \\
 & (masked) & GLU & 12.62 & 29.22 \\
 & & GAU & \textbf{8.35} & 23.07 \\
 \cmidrule{2-5}
 & \multirow{5}{*}{Weight Norm} & ReLU & 9.98 & 24.87 \\
 & & cReLU & 11.25 & 26.87 \\
 & & lReLU & 11.87 & 27.54 \\
 & & GLU & 11.05 & 27.10 \\
 & & GAU & 11.25 & 27.70 \\
 \midrule
 \multirow{4}{*}{Jasper 10x4} & Batch Norm & ReLU & \textbf{6.15} & \textbf{17.58} \\
 \cmidrule{2-5}
 & Layer Norm & ReLU & 6.56 & 18.48 \\
 & (Masked) & GAU & 7.14 & 19.19 \\
 \bottomrule
\end{tabular}
}
\end{table}

\begin{table}[th]
\centering
\caption{Sequence Masking: Greedy WER, LibriSpeech for Jasper 10x4 after 50 epochs}
\label{tab:SeqMask}
\scalebox{0.8}{
\begin{tabular}{c c c c} 
 \toprule
 \multirow{2}{*}{\textbf{Model}} & \multirow{2}{*}{\textbf{Masking}} & \multicolumn{2}{c}{\textbf{Dev}} \\
 & & \textbf{Clean} & \textbf{Other} \\
 \midrule
 Jasper DR 10x4 & None & 5.88 & 17.62 \\
 Jasper DR 10x4 & BN Mask & 5.92 & 17.63 \\
 Jasper DR 10x4 & Conv Mask & \textbf{5.66} & \textbf{16.77} \\
 Jasper DR 10x4 & Conv+BN Mask & 5.80 & 16.97 \\
 \bottomrule
\end{tabular}
}
\end{table}

\subsection{Residual Connections}
For models deeper than Jasper 5x3, we observe consistently that residual connections are necessary for training to converge. In addition to the simple residual and dense residual model described above, we investigated DenseNet \cite{huang2016} and DenseRNet \cite{Tang2018} variants of Jasper. Both connect the outputs of each sub-block to the inputs of following sub-blocks within a block. DenseRNet, similar to Dense Residual, connects the output of each block to the input of all following blocks. DenseNet and DenseRNet combine residual connections using concatenation whereas Residual and Dense Residual use addition. We found that Dense Residual and DenseRNet perform similarly with each performing better on specific subsets of LibriSpeech. We decided to use Dense Residual for subsequent experiments. The main reason is that due to concatenation, the growth factor for DenseNet and DenseRNet requires tuning for deeper models whereas Dense Residual does not have a growth factor.

\begin{table}[th]
\centering
\caption{Residual Connections: Greedy WER, LibriSpeech for Jasper 10x3 after 400 epochs.  All models sized to have roughly the same parameter count.}
\label{tab:Res}
\scalebox{0.8}{
\begin{tabular}{c c c c} 
 \toprule
 \multirow{2}{*}{\textbf{Model}} & \multirow{2}{*}{\textbf{\#params, M}} & \multicolumn{2}{c}{\textbf{Dev}} \\
 &  & \textbf{Clean} & \textbf{Other} \\
 \midrule
 Residual & 201 & 4.65 & 14.36 \\
 Dense Residual & 211 & 4.51 & \textbf{14.15} \\
 DenseNet & 205 & 4.77 & 14.55 \\
 DenseRNet & 211 & \textbf{4.32} & 14.21 \\
 \bottomrule
\end{tabular}
}
\end{table}

\subsection{Language Model}
A language model (LM) is a probability distribution over arbitrary symbol sequences $P(w_1, ..., w_n)$ such that more likely sequences are assigned higher probabilities. LMs are frequently used to condition beam search. During decoding, candidates are evaluated using both acoustic scores and LM scores. Traditional N-gram LMs have been augmented with neural LMs in recent work \cite{zeyer2018improved, Povey2018SemiOrthogonalLM, CAPIO2017}.

\begin{figure}[t!h]
  \centering
  \includegraphics[width=\linewidth]{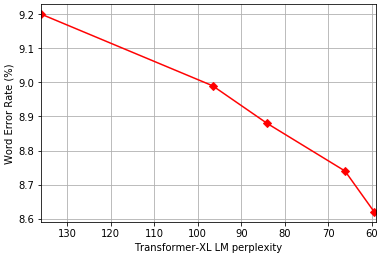}
  \caption{LM perplexity vs WER. LibriSpeech dev-other.  Varying perplexity is achieved by taking earlier or later snapshots during training.}
  \label{fig:wer_vs_pplx}
\end{figure}

We experiment with statistical N-gram language models \cite{heafield2011kenlm} and neural Transformer-XL \cite{dai2018transformer} models. Our best results use acoustic and word-level N-gram language models to generate a candidate list using beam search with a width of 2048. Next, an external Transformer-XL LM rescores the final list. All LMs were trained on datasets independently from acoustic models. We show results with the neural LM in our Results section. We observed a strong correlation between the quality of the neural LM (measured by perplexity) and WER as shown in Figure \ref{fig:wer_vs_pplx}. 
\iffalse{
\begin{table}[th]
\centering
\caption{Effects of LM on WER, LibriSpeech and WSJ. ``No LM" uses greedy decoding. A beam size of 2048 was used for n-gram and Transformer-XL.}
\label{tab:LibriLMRes}
\scalebox{0.8}{
\begin{tabular}{c c c c c c} 
 \toprule
 \multirow{2}{*}{\textbf{LM}} & \multicolumn{2}{c}{\textbf{LibriSpeech test}} & \multicolumn{2}{c}{\textbf{WSJ}} \\
 & \textbf{Clean} & \textbf{Other} & \textbf{nov93} & \textbf{nov92} \\
 \midrule
 No LM & 3.86 & 11.95 & 16.13 & 13.34 \\
 6-gram KenLM & 3.34 & 9.68 & 9.91 & 7.10 \\
 Transformer-XL & \textbf{2.95} & \textbf{8.82} & \textbf{9.29} & \textbf{6.86} \\
 \bottomrule
\end{tabular}
}
\end{table}
}\fi

\begin{table*}[!ht]
\centering
\caption{LibriSpeech, WER (\%)}
\label{tab:LibriSpeechResults}
\scalebox{0.8}{
\begin{tabular}{c c c c c c c} 
 \toprule
 {\textbf{Model}} & \textbf{E2E} & {\textbf{LM}} & {\textbf{dev-clean}} & {\textbf{dev-other}} & {\textbf{test-clean}} & {\textbf{test-other}} \\
 \midrule
 CAPIO (single) \cite{CAPIO2017} & N & RNN & 3.02 & 8.28 & 3.56 & 8.58 \\
 pFSMN-Chain \cite{Yang2018} & N & RNN & 2.56 & 7.47 & 2.97 & \textbf{7.5} \\
 \midrule
 DeepSpeech2 \cite{DeepSpeech2}  & Y & 5-gram & - & - & 5.33 & 13.25 \\
 Deep bLSTM w/ attention \cite{zeyer2018improved} & Y & LSTM & 3.54 & 11.52 & 3.82 & 12.76 \\
 wav2letter++ \cite{zeghidour2018} & Y & ConvLM & 3.16 & 10.05 & 3.44 & 11.24 \\
 LAS + SpecAugment \footnotemark \phantom{} \cite{park2019} & Y & RNN & - & - & \textit{2.5} & \textit{5.8} \\
 
 \midrule
 Jasper DR 10x5 & Y & -      & 3.64 & 11.89 & 3.86 & 11.95 \\ 
 Jasper DR 10x5 & Y & 6-gram & 2.89 &  9.53 & 3.34 & 9.62 \\ 
 Jasper DR 10x5 & Y & Transformer-XL & 2.68 & 8.62 & \textbf{2.95} & 8.79 \\ 
 Jasper DR 10x5 + Time/Freq Masks \footnotemark[\value{footnote}] & Y & Transformer-XL & 2.62 & 7.61 & 2.84 & 7.84 \\ 
 \bottomrule
\end{tabular}
}
\end{table*}

\subsection {NovoGrad}
For training, we use either Stochastic Gradient Descent (SGD) with momentum or our own \textit{NovoGrad}, an optimizer similar to Adam \cite{kingma}, except that its second moments are computed per layer instead of per weight. Compared to Adam, it reduces memory consumption and we find it to be more numerically stable.

At each step $t$, NovoGrad computes the stochastic gradient $g^l_t$ following the regular forward-backward pass. Then the second-order moment $v^l_t$ is computed for each layer $l$ similar to ND-Adam \cite{zhang2017ndadam}:
\begin{equation}
    v^l_t = \beta_2 \cdot v^l_{t-1} + (1-\beta_2) \cdot ||g^l_t||^2
\end{equation}
The second-order moment $v^l_t$ is used to re-scale gradients $g^l_t$ before calculating the first-order moment $m^l_t$:
\begin{equation}
    m^l_t = \beta_1 \cdot m^l_{t-1} + \frac{g^l_t}{\sqrt{v^l_t+\epsilon}}
\end{equation}
If L2-regularization is used, a weight decay $d \cdot w_{t}$ is added to the re-scaled gradient (as in AdamW  \cite{loshchilov2018}): 
\begin{equation}
    m^l_t = \beta_1 \cdot m^l_{t-1} + \frac{g^l_t}{\sqrt{v^l_t + \epsilon}} + d \cdot w_{t}
\end{equation}
Finally, new weights are computed using the learning rate $\alpha_t$:
\begin{equation}
    w_{t+1} = w_{t} -\alpha_t \cdot m_t 
\end{equation}
Using NovoGrad instead of SGD with momentum, we decreased the WER on dev-clean LibriSpeech from 4.00\% to 3.64\%, a relative improvement of 9\% for Jasper DR 10x5. For more details and experiment results with NovoGrad, see \cite{novograd2019}.
%}
%\fi

\iffalse{
\section{Discussion}
Our works builds on the very similar principles as two leading end-to-end neural acoustic models: Listen-Attend-Spell (LAS) v2 \cite{Zhang2017}, and wav2letter++ \cite{zeghidour2018}. 

LAS models are attention-based NN models, based on the sequence to sequence approach which don't use CTC loss function. LAS v2 applied batch norm, ReLU, similar to Jasper, but LAS main building block is Convolutional LSTM, where inner inner products are replaced with convolutions over previous state. Note that this is completely different from Jasper 1D-convolutions which operates over time. To increase model depth, authors added 2 convolutional layers before LSTM block, similar to DeepSpeech2 model. They also added identity residual connections, since for LSTM cells residual projections didn't help much. Training and inference for Jasper are faster than for LAS since LSTM-based models  are inherently sequential. 

Jasper follows wav2letter and wav2letter++, which are both fully convolutional acoustic NN models. These are main architecture differences between wav2letter* and Jasper: we replaced 1) weight normalization with batch normalization 2) gated linear units with ReLU\. These two changes made training faster and more robust. Using ReLU and batch norm open path to fast inference since these two layers can be fused with convolutional layer. Thanks to using residual connections we can train much deeper model than wav2letter++, which has only 19 layers. 
}
\fi

\footnotetext{We include the latest SOTA which was achieved by Park et al. \cite{park2019} after our initial submission. We add results for Jasper with time and frequency masks similar to SpecAugment. We use 1 continuous time mask of size $T \sim U(0, 99)$ time steps, and 1 continuous frequency mask of size $F\sim U(0, 26)$ frequency bands.}

\section{Results}
We evaluate Jasper across a number of datasets in various domains. In all experiments, we use dropout and weight decay as regularization. At training time, we use 3-fold speed perturbation with fixed +/-10\% \cite{Ko2015} for LibriSpeech. For WSJ and Hub5'00, we use a random speed perturbation factor between [-10\%, 10\%] as each utterance is fed into the model. All models have been trained on NVIDIA DGX-1 in mixed precision \cite{micikevicius2017mixed} using OpenSeq2Seq \cite{OpenSeq2Seq}. Pretrained models and training configurations are available from ``https://nvidia.github.io/OpenSeq2Seq/html/speech-recognition.html''.

\subsection{Read Speech}

We evaluated the performance of Jasper on two read speech datasets: LibriSpeech and Wall Street Journal (WSJ). For LibriSpeech, we trained Jasper DR 10x5 using our NovoGrad optimizer for 400 epochs. We achieve SOTA performance on the test-clean subset and SOTA among end-to-end speech recognition models on test-other.

We trained a smaller Jasper 10x3 model using the SGD with momentum optimizer for 400 epochs on a combined WSJ dataset (80 hours): LDC93S6A (WSJ0) and LDC94S13A (WSJ1). The results are provided in Table \ref{tab:WSJResults}.

\begin{table}[!h]
\centering
\caption{WSJ End-to-End Models, WER (\%)}
\label{tab:WSJResults}
\scalebox{0.8}{
\begin{tabular}{c c c c c c c} 
 \toprule
 {\textbf{Model}} & {\textbf{LM}} & {\textbf{nov93}} & {\textbf{nov92}} \\
  \midrule
 seq2seq + deep conv \cite{Zhang2017} & - & - & 10.5 \\
 wav2letter++ \cite{zeghidour2018} & 4-gram & 9.5 & 5.6 \\
 wav2letter++ \cite{zeghidour2018} & ConvLM & 7.5 & 4.1 \\
 E2E LF-MMI \cite{Hadian2018} & 3-gram & - & 4.1 \\
  \midrule
 Jasper 10x3 & - & 16.1 & 13.3 \\ 
 Jasper 10x3 & 4-gram & 9.9 & 7.1 \\ 
 Jasper 10x3 & Transformer-XL & 9.3 & 6.9 \\ 
 \bottomrule
\end{tabular}
}
\end{table}
\subsection{Conversational Speech}
We also evaluate the Jasper model's performance on a conversational English corpus. The Hub5 Year 2000 (Hub5'00) evaluation (LDC2002S09, LDC2002T43) is widely used in academia. It is divided into two subsets: Switchboard (SWB) and Callhome (CHM). The training data for both the acoustic and language models consisted of the 2000hr Fisher+Switchboard training data (LDC2004S13, LDC2005S13, LDC97S62). Jasper DR 10x5 was trained using SGD with momentum for 50 epochs. We compare to other models trained using the same data and report Hub5'00 results in Table \ref{tab:Hub5Results}.

\begin{table}[!h]
\centering
\caption{Hub5'00, WER (\%)}
\label{tab:Hub5Results}
\scalebox{0.8}{
\begin{tabular}{c c c c c} 
 \toprule
 {\textbf{Model}} & \textbf{E2E} &\textbf{LM} & {\textbf{SWB}} & {\textbf{CHM}} \\
 \midrule
 LF-MMI \cite{Hadian2018} & N & RNN & 7.3 & 14.2 \\
 \midrule
 Attention Seq2Seq \cite{Weng2018} & Y & - & 8.3 & 15.5 \\
 RNN-T \cite{Battenberg2017} & Y & 4-gram & 8.1 & 17.5 \\
 Char E2E LF-MMI \cite{Hadian2018} & Y & RNN & 8.0 & 17.6 \\
 Phone E2E LF-MMI \cite{Hadian2018} & Y & RNN & 7.5 & 14.6 \\
 CTC + Gram-CTC & Y & N-gram & 7.3 & 14.7 \\
 \midrule
 Jasper DR 10x5 & Y & 4-gram & 8.3 & 19.3 \\
 Jasper DR 10x5 & Y & Transformer-XL & 7.8 & 16.2 \\ 
\bottomrule
\end{tabular}
}
\end{table}

\iffalse{
\begin{table}[!h]
\scalebox{0.8}{
\begin{threeparttable}
\centering
\caption{Hub5, WER (\%)}
\label{tab:Hub5Results}
\begin{tabular}{l c c c c} 
 \toprule
 {\textbf{Model}} & \textbf{\thead{AM\\Data}} & \textbf{\thead{LM\\Data}} & {\textbf{SWB}} & {\textbf{CHM}} \\
 \midrule
 CAPIO (fusion) \cite{CAPIO2017} & F+S+C & F+S+C+O & 5.0 & 9.1 \\
 Karuta, et. al (fusion) \cite{Kurata2017LanguageMW, Saon2017} & F+S+C & F+S+C+O & 5.1 & 9.9 \\ 
 CAPIO (single) \cite{CAPIO2017} & F+S+C & F+S+C+O & 5.6 & 10.5 \\
 \midrule
 Hannun, et. al. \cite{DeepSpeech2014} & F+S & F+S & 12.6 & 16.0 \\
 LF-MMI \cite{Povey+2016} & F+S & F+S & 8.5 & 15.3 \\
\midrule
 Jasper DR 10x5 (4-gram) & F+S & F+S & 8.3 & 16.0 \\
 Jasper DR 10x5 (TF-XL) & F+S & F+S & 7.8 & 16.2 \\ 
\bottomrule
\end{tabular}
\begin{tablenotes}
\item F: Fisher Corpus, S: Switchboard Corpus
\item C: Callhome Corpus, O: Other Data
\end{tablenotes}
\end{threeparttable}
}
\end{table}
}\fi

We obtain good results for SWB. However, there is work to be done to improve WER on harder tasks such as CHM.

\section{Conclusions}
We have presented a new family of neural architectures for end-to-end speech recognition. Inspired by wav2letter's convolutional approach, we build a deep and scalable model, which requires a well-designed residual topology, effective regularization, and a strong optimizer. As our architecture studies demonstrated, a combination of standard components leads to SOTA results on LibriSpeech and competitive results on other benchmarks. Our Jasper architecture is highly efficient for training and inference, and serves as a good baseline approach on top of which to explore more sophisticated regularization, data augmentation, loss functions, language models, and optimization strategies. We are interested to see if our approach can continue to scale to deeper models and larger datasets.

\bibliography{Jasperbib}

\end{document}